\documentclass[manuscript]{aastex}

\slugcomment{}

\shorttitle{The Distance to NGC 4258 and the Hubble Constant}
\shortauthors{Humphreys et al.}

\begin{document}

\title{Toward a New Geometric Distance to the Active Galaxy NGC 4258. III. Final Results and the Hubble Constant}

\author{E. M. L. Humphreys\altaffilmark{1,2}}

\author{M. J. Reid\altaffilmark{2}}

\author{J. M. Moran\altaffilmark{2}}

\author{L. J. Greenhill\altaffilmark{2}}

\author{A. L. Argon\altaffilmark{2}}

\affil{$^1$European Southern Observatory,
                 Karl-Schwarzschild-Str. 2,
                 85748 Garching bei M\"{u}nchen, Germany: ehumphre@eso.org}
\affil{$^2$Harvard-Smithsonian Center for Astrophysics,
                 60 Garden Street, Cambridge, MA 02138, USA}

\begin{abstract}
We report a new geometric maser distance estimate to the active galaxy
NGC 4258. The data for the new model are maser
line-of-sight velocities and sky positions from 18 epochs
of Very Long Baseline Interferometry observations, and line-of-sight 
accelerations
measured from a 10-year monitoring program of the 22~GHz maser emission
of NGC 4258. The new model includes both disk warping and confocal elliptical 
maser orbits with differential precession. 
The distance to NGC 4258 is 7.60 $\pm$0.17$\pm$0.15 Mpc, 
  a 3\% uncertainty including formal fitting and systematic terms.
The resulting Hubble Constant,
 based on the use of the Cepheid Variables in NGC 4258 to recalibrate the Cepheid distance scale 
\citep{Riess2011}, is H$_{0}$ = 72.0 $\pm$ 3.0 km s$^{-1}$ Mpc$^{-1}$.
\end{abstract}

\keywords{Cosmology: distance scale --- Galaxies: individual:NGC 4258 --- Galaxies: nuclei --- Masers --- Techniques: interferometric}

\section{Introduction}
\label{s:intro}

Observations and modeling of masers in the circumnuclear disk
of the Seyfert 2/LINER galaxy NGC 4258 have resulted in a
distance estimate of 7.2 $\pm$ 0.2 (random) $\pm$ 0.5
(systematic) Mpc \citep[][hereafter H99]{Herrnstein1999},
in which the systematic component largely allowed for
the potential effects of unmodeled eccentric orbits.
 The goal of the current work
is to reduce this 
uncertainty 
(\citealt[][hereafter Paper I; ]{Humphreys2005,Argon2007}\citealt[][hereafter Paper II]{Moran2007,Humphreys2008}).
In this paper, we report a new distance estimate for
NGC 4258 in which considerably more data have been used:
18 epochs of Very Long Baseline Interferometry (VLBI) data compared
with the 4 epochs used in H99.
 Also,  significant progress has been made in the modeling approach, 
 including the possibility of eccentricity in the maser orbits.

NGC 4258 cannot be used to determine the Hubble Constant H$_{0}$ directly to high accuracy, 
since the galaxy is 
relatively close and its peculiar motion could be a large fraction of its redshift.
However, it can be used as an anchor for the Cepheid-calibrated extragalactic distance scale,
 in addition to the Large Magellanic Cloud and the Milky Way, to reduce uncertainty in  H$_{0}$.
The Hubble Space Telescope Key Project measured H$_{0}$=72 $\pm$3 $\pm$7  km s$^{-1}$ Mpc$^{-1}$ \citep{Freedman2001}.
Using the current maser distance to NGC 4258 of H99, \citet{Macri2006} 
recalibrated the Cepheid period-luminosity relation to obtain  H$_{0}$=74 $\pm$3 $\pm$6 km s$^{-1}$ Mpc$^{-1}$.
\citet{Riess2011,Riess2012} also attempted to recalibrate the Cepheid relation using
an unpublished preliminary maser distance to NGC 4258 of 7.28 Mpc $\pm$ 3\%, obtaining 
H$_{0,4258}$=74.8 $\pm$  3.1 km s$^{-1}$ Mpc$^{-1}$, a fractional accuracy of 4\%,  where H$_{0,4258}$
is the Hubble Constant determined when the sole anchor galaxy is NGC 4258.

We describe the input data for the distance models
in Section~\ref{s:data},     the models themselves in Section~\ref{s:models}, 
and we compare with the approach of previous work in Section~\ref{s:compare}.  
We present the results in Section~\ref{s:results}, discuss the impact
of the new maser distance on H$_{0}$ in Section~\ref{s:discuss} and
summarize conclusions in Section~\ref{s:conclu}.

\section{Input Data}
\label{s:data}

The data for our maser geometric distance measurement comes 
from VLBI mapping to obtain maser sky positions, augmented by single dish monitoring of spectra to measure centripetal accelerations. 
The data used to determine the maser disk geometry and the distance
to NGC 4258 consist of maser emission positions ($X$,$Y$),
line-of-sight (LOS) velocities ($v_{los}$), and LOS accelerations
($a_{los}$). 
We measured position and velocity data at 18 epochs using
  VLBI with the methods described in Paper I.  We also estimated
  accelerations (from time varying Doppler shifts) from spectra
  obtained during the VLBI observations, supplemented by spectra
  from the Jansky Very Large Array (VLA\footnotemark) and the Robert C. Byrd
  Green Bank Telescope (GBT).  The acceleration estimates
  were documented in Paper II.
The resulting data set consisted of $\sim$10,000 data points.
In order to create a more tractable dataset to use in the disk fitting
programs described here, we binned the data acquired at different epochs in velocity
(using a bin width of 1 km s$^{-1}$) yielding the reduced
   data set described in Table~\ref{t:inputdata}.

The entries in Table~\ref{t:inputdata} are listed separately for the high-velocity 
red-shifted and blue-shifted maser emission occurring at about  $v_{sys}$$\pm$1000 km~s$^{-1}$ 
respectively, where  $v_{sys}$ is the galactic systemic velocity, and for the low-velocity 
(systemic) maser emission ocurring at about $v_{sys}$. We give the range of LOS velocities
measured for maser emission over all epochs of the observations, and the associated
ranges of maser $X$ and $Y$ sky positions from VLBI observations. The LOS 
accelerations for the maser emission were determined using both single-dish and interferometric data   
using a Gaussian decomposition method that simultaneously fit Gaussians to maser spectra at
multiple epochs to determine drifts in velocity over time. The number of data points in the reduced dataset is also provided
separately for the high-velocity and low-velocity emission in Table~\ref{t:inputdata}.   

\footnotetext{The National Radio Astronomy Observatory is a facility of the National Science Foundation operated under cooperative agreement by Associated Universities, Inc.}

\section{3-D Disk Fitting Model}
\label{s:models}

\subsection{Overview of the Model}

We employed a Bayesian 3-D global disk fitting program
\citep{Reid2013} in order to determine the maser disk geometry
and estimate distance (Sect.~\ref{ss:bayesian}). 
Our model has 13 global
 parameters, which we describe in the following paragraphs.

The global disk parameters include distance ($D$), black hole mass
($M_{bh}$), galaxy systemic velocity ($v_{0}$), black hole x-position
($X_{0}$) and y-position ($Y_{0}$).

A simple model for elliptical maser orbits is one
in which eccentric orbits precess coherently in the disk, such
that the orbits are aligned and nested \citep{Statler2001,Statler2002}.
This scenario could be described by a single eccentricity ($e$)
and periapsis angle ($\omega$). However, it may not be a
valid description for the nuclear maser disk, since  zero viscosity
would be required and there is no mechanism by which such a disk would
accrete.
Therefore the model we investigated is one for a viscous disk in which
orbits undergo differential precession and the periapsis angle is
described by a leading/trailing spiral \citep{Armitage2008}.
This introduces a radial gradient in periapsis angle $\omega$, such
that $\omega_r = \omega_0 + r d\omega / dr$, where the reference
angular radius for $\omega_0$ is $r = 0$.

We modeled the warped disk as a surface whose position angles and
inclinations vary as smooth functions of radius ($\Omega_r,i_r$).
\citet{Herrnstein2005} found that the warp is well described
by an inclination warp of $i_r = i_0 + r di/dr$
and a position angle warp of
$\Omega_r = \Omega_0 + r d\Omega/dr + r^2 d^2\Omega/dr^2$, 
 where in this paper the reference
radius for $i_0$ and $\Omega_0$ is $r = 0$. However, we investigated the effect
on distance of inclusion of a second-order inclination
warp term as part of our quantification of systematic uncertainties
in Sect.~\ref{s:systematics}.
The disk inclination angle is
measured between the observer's LOS and the negative spin
axis of the maser disk; position angle is measured East of North.

In addition to the 13 global disk parameters, the angular radius ($r$) and
azimuth angle ($\phi$) in the disk of each maser spot
were included as parameters. In total, there were 753 parameters in the
model, which we fitted to 1262 data points.

The geometry of the sky and disk coordinate systems, and the derivation of the equations
used in the modeling, are described in Appendix~\ref{a:coordinates}. In the
modeling, we assumed that masers orbit a 
point mass located at the focus of confocal ellipses of common eccentricity.

The model LOS velocities for masers (Appendix~\ref{a:velocity}) are 
given by Keplerian rotation, for which

\begin{equation}
v_{los,model}=v_r \sin i_r \sin \phi  + v_\gamma \sin i_r \cos \phi  + v_0, 
\end{equation}

\noindent where $v_r$ and $v_\gamma$ are the radial and tangential components given by $v_{r}  =  [GM_{bh}/(r D (1+e\cos\gamma))]^{1/2} e\sin\gamma$
and $v_{\gamma}  = [GM_{bh}(1+e\cos\gamma)/(r D)]^{1/2}$, where $\gamma=\phi - \omega_r$ is the angle between the maser and perihelion.
However, due to significant special and general relativistic effects in the
transformation of observed frequency to LOS velocity for the masers,
we used relativity-corrected LOS velocities $v'_{los,model}$ (Appendix~\ref{a:relativity}).

The component of centripetal acceleration in the LOS is given by

\begin{equation}
a_{los,model} =  \frac{-GM_{bh}}{(rD)^2}\sin i_r\sin\phi , 
\end{equation}

\noindent and sky positions of masers are given by

\begin{eqnarray}
X & = & r(\sin\Omega_r\cos\phi - \cos\Omega_r\cos i \sin\phi) + X_0\nonumber\\
Y & = & r(\cos\Omega_r\cos\phi + \sin\Omega_r\cos i \sin\phi) + Y_0.
\end{eqnarray}

\noindent where ($X_0,Y_0$) is the sky position of the disk dynamical center measured relative to maser emission at 510 km s$^{-1}$. It is important to note
that this reference position in the maser disk is
defined by  velocity, not by a specific clump sometimes called a maser ``spot''
or ``feature.''

In the modeling, we compared observed maser ($X_{obs}, Y_{obs}, v'_{obs}, a_{obs}$) with
model values ($X_{model}$,$Y_{model}$,$v'_{los,model}$,$a_{los,model}$) to determine
distance. Essentially,  the rotation curve of the high-velocity maser emission 
constrains ${\cal M}^{1/2} \sin i_r$ where ${\cal M}=(M_{bh}/D)$ and the accelerations of the systemic maser features 
provide distance via $D = (-G {\cal M} / r^2 a_{los,model})  \sin i_r \sin\phi$.

\subsection{The Bayesian Fitting Program
 \label{ss:bayesian}}

We used a Markov chain Monte Carlo (MCMC) fitting program in which the Metropolis-Hastings algorithm was used to select the Markov-chain 
trial parameters. 
We used an initial  2,000,000 ``burn-in'' trials to be discarded at the start of each
fitting run and then saved the outcome of the subsequent 10,000,000 trials.
The MCMC parameter step size was adjusted every 100,000 steps during burn-in in order to 
maintain an acceptance rate of 25\% of the trial parameters. 
The trial values for each parameter were binned,
yielding marginalized posteriori probability distributions,
from which we quote median values and 68\% confidence
(``1$\sigma$") ranges.

The MCMC method is not designed to yield the ``best" minimum in $\chi^2$ space.
 However we output the lowest $\chi^2$ of the MCMC trials, calculated using

\begin{eqnarray}
\chi^2 & = & \sum_n \left(\frac{v'_{obs} - v'_{los,model}}{\sigma_v}\right)^2 + \left(\frac{X_{obs} - X_{model}}{\sigma_x}\right)^2 \nonumber \\
       &   & + \left(\frac{Y_{obs} - Y_{model}}{\sigma_y}\right)^2 + \left(\frac{a_{obs} - a_{los,model}}{\sigma_a}\right)^2
\end{eqnarray}

\noindent and scaled the 1$\sigma$ uncertainties for
each global disk parameter by $\sqrt{\chi^2/N}$, where $N$ is the number of degrees of freedom.

\section{Comparison with Previous Approaches}
\label{s:compare}

There are several notable differences between the current work
and that of H99.
First, in H99,  the distance calculation was performed in two steps.
A warped disk model was determined without the incorporation of
systemic feature accelerations. Distance was then calculated
in a decoupled second step involving these accelerations. 
In the current work, we fitted the data in a
single step, simultaneously adjusting all parameters.
The second difference is that, in H99,
$r$ and $\phi$ were not treated as model parameters with associated
uncertainties. In their model,
allowed loci of ($r$,$\phi$) were determined from line-of-sight velocities, 
and $r$ and $\phi$ were assumed to be perfectly determined. In our
work, maser emission $r$ and $\phi$ are included as parameters in the
models. 
The third difference is that orbits were fixed to be circular in H99,
whereas in the current model, eccentric maser orbits are included.

Another difference between this work and the investigation of 
disk warping performed by \citet{Herrnstein2005}
is that they performed a fit to the high-velocity maser emission
only. In this work, we fit to all observables for both
the low-velocity (systemic) and the high-velocity emission.

Finally, the distance estimate of H99 was based on 4 VLBI epochs, 
whereas we analyze 18 VLBI epochs. The acceleration dataset of
H99 included accelerations for fewer than 20 systemic maser features 
measured from 4 epochs over 3 years. In the
current work, we included acceleration measurements for
both systemic and high-velocity features from a monitoring program
lasting up to $\sim$10 years, described in detail in Paper II of this
series.

\section{Maser Distance to NGC 4258}
\label{s:results}

\subsection{The ``Base'' Model}
\label{s:basemodel}

We used the MCMC program to establish an optimum or ``base'' model for the NGC 4258
maser disk geometry and distance. 
The initial global disk parameters to the fitting program were taken to be approximately
those given by \citet[][Table~\ref{t:jrhparams}]{Herrnstein2005}. 
In order to ensure that
eccentricity space was fully explored, we set initial values of
eccentricity as high as 0.5.
All global disk parameters had flat priors. 
Input values for the ($r$,$\phi$) maser feature polar disk coordinate parameters were estimated 
using observational data and assuming Keplerian rotation. These parameters were assigned the following loose
prior uncertainties: in the range $\pm$ $\sim$1--2 mas  and
$\pm$20$^{\rm o}$ for the high-velocity ($r$,$\phi$). Systemic features had prior uncertainties of $\pm$2 mas and $\pm$1$^{\rm o}$.

Inspection of data among different observing epochs showed scatter
larger than expected from the formal uncertainties. In order to account
for this, error floors were added to the formal uncertainties in
quadrature.
The error floors effectively
weight data differently in the MCMC modeling, which is somewhat subjective, but
which is investigated as part of our analysis of systematic uncertainties in the
modeling.
The error floors used here
were  0.02 and 0.03 mas for maser $x$ and $y$ sky positions respectively
(VLBI position uncertainties are larger in the $y$-data due to north-south beam elongation of a factor of 1.5 for NGC 4258
at a declination of $+$47$^{\rm o}$), 1.0 km s$^{-1}$ for all maser feature
velocity error floors, and 0.3 km s$^{-1}$ yr$^{-1}$ for acceleration data.
The impact of selection of error floors on distance is investigated in Sect.~\ref{s:systematics} as part of our quantification of systematic uncertainties. 
We note that it is important that the high-velocity maser features have sufficient weight so as
to constrain the rotation curve,  since it provides the strongest constraints on $M_{bh}/D$.

The resulting base model (Table~\ref{t:basemodel}; Figures~1--7)
has parameters for the maser disk warping that are in good agreement with those
found in the fit to high-velocity maser features only by  \citet{Herrnstein2005}.
The eccentricity of maser orbits is low at 0.006$\pm$0.001. We estimate
a maser distance of 7.6$\pm$0.17 Mpc  (formal fitting error scaled by $\sqrt{\chi^2/N}$ where $\chi^2/N=1.4$).

\subsection{Investigation of Systematic Uncertainties}
\label{s:systematics}

In addition to the formal uncertainty derived directly
from the MCMC fitting program, we investigated various sources
of systematic uncertainty that could affect the maser distance to NGC 4258.
Firstly, we varied the values of the error floors that we added in
quadrature to the formal errors on our dataset to assess
their impact on distance.  
Secondly, we investigated the effect of changing the initial parameter
values and the random number generator seed used for selecting MCMC
trial disk parameters.  
In particular we
ran the MCMC trials for a wide range of input distances (7.1--8.2 Mpc)
to make sure the that no memory of the starting value was retained.
Thirdly, we tested the impact on distance if we allowed
a second order inclination warp term and what would happen if
we constrained the maser orbits to be circular.  The results
of these tests are presented in Table~\ref{t:uncertainties}.

In Table~\ref{t:uncertainties}, we found that the largest 
sources of systematic uncertainty in the distance estimation are
given by starting the MCMC code with different initial values
for distance which contributes 1.5\%, and the possibility of unmodeled spiral
structure (Section~\ref{sss:spirals}) which
contributes $\sim$1\%. Other sources of systematic
uncertainty were all found to cause $<$1\% change in
distance. Summing the uncertainty terms in quadrature
gives a total systematic uncertainty of $\pm$ 0.15 Mpc or $\pm$ 2\%.

\subsubsection{Systematic Uncertainty Due to Unmodeled Spiral Structure}
\label{sss:spirals}

In Paper II, we showed evidence for periodic structure in the NGC 4258
high-velocity maser emission feature position distribution
\citep[originally noted by][]{Maoz1995}. We also found a persistent slope in the 
line-of-sight accelerations of low-velocity maser emission as a function
of disk impact parameter. The slope was first seen by \citet{Greenhill1995a} 
and so has persisted for at least 6 to 7 years. This is notable because if it were just an ``accidental" quirk in the radial distribution of systemic maser features, it should have 
systematically moved or rotated out of the line of sight within in a few years.
 However, both of these results could be explained by the
 presence of spiral structure in the disk.

Stability arguments support 
self-gravitating structure formation, at least in the outer disk regions.
The Toomre $Q$-parameter 
calculated
for $n(H_{2})$=10$^{10}$ cm$^{-3}$, a disk full height of 12 $\mu$as
 (Paper I), and a central black hole mass of 4.0 $\times$ 10$^7$ 
M$_\odot$ varies between 10 to 1, respectively, for the maser disk between 0.1 and
0.30 pc (but we note that $Q$ would be between 26 to 3 if the velocity dispersion is used in
place of the sound speed). Although $Q$ $<$ 1 is required for instability
to axisymmetric perturbations, instability to non-axisymmetric perturbations
occurs in the 1 $<$ $Q$ $<$ 2 regime and spiral structure could form.
In Paper II, we performed ``proof-of-concept'' $N$-body modeling that
showed that the gradient in the low velocity acceleration data
could be reproduced by a spiral arm of mass 15\% of the upper limit mass
for the maser disk given by \citet{Herrnstein2005}.

To assess the impact of unmodeled spiral structure on a distance
estimate, we took the same simulated data that reproduced the gradient
in acceleration data in Paper II and fed it into our fitting program
to see how well it could retrieve the distance. We found that
the presence of the spiral arm manifested itself as a significant shift in the recovered
black hole $x_{0}$ position by $\sim$0.2 mas,  the galaxy systemic velocity by $\sim$40 km s$^{-1}$, 
eccentricity by $\sim$0.08, and the distance 
by 3.6 \% (Table~\ref{t:spiralstructure}).

We know that such effects cannot be ``hiding'' in our data and modeling. In our
modeling we obtain a systemic velocity of 474.3$\pm$0.5 km s$^{-1}$, which is very close to that independently
obtained by \citet{Cecil1992} of 472$\pm$4 km s$^{-1}$ for H$\alpha$ observations using the
Hawaii Imaging Fabry-Perot Interferometer. In addition, we find a much lower eccentricity
for the NGC 4258 disk at 0.006$\pm$0.001 compared with the eccentricity of 0.08  found for the
simulated data that included spiral structure. 

However, it is possible that a smaller effect, at about the 10 to 20\% level of that obtained
for our spiral structure example (i.e., a change in systemic velocity of about 10 km s$^{-1}$ or an eccentricity of 
about 0.01 to 0.02) could perhaps be relevant to the NGC 4258 data set. We therefore adopt a conservative 1\% uncertainty in distance 
due to unmodeled spiral structure and await further study of spiral structure in AGN nuclear disks  
to clarify the best way to assess this term in future work.

\subsection{Parameter Correlations}
\label{ss:correlation}

The global disk parameter correlation matrix for the base model is given in Table~\ref{t:correlations}.
The correlation coefficient of $D$ and $M_{bh}$ of 0.998 superficially suggests that these parameters are degenerate
and calls into question whether these values can be determined reliably from the modeling. 
 However, this naturally occurs when combining constraints on $M_{bh}$ and $D$
   from different types of data that are sensitive to different
   powers of these parameters.  In such a case,
   while $M_{bh}$ and $D$ are highly correlated over a range of values,
   their probable values are also strongly bounded.

In order to investigate how $M_{bh}$ and $D$ are constrained, consider an edge-on disk 
with systemic maser features at a fixed 
angular radius $r_{sys}$ and high-velocity maser features that lie on
the disk midline and at disk angular radii of  $r_{hv}$.
We can then formulate three
mass---distance relations. The first is from the high-velocity
feature Keplerian rotation curve and is $M = G M_{bh} = WD$, 
where $W = v^2_{hv} r_{hv} $. The second is from the slope
of the systemic feature position-velocity diagram, with
slope $s = M^{0.5} r_{sys}^{-0.5} D$. 
The third is that the line-of-sight component of accelerations of
the systemic features is given by $a = M r_{sys}^{-2}$, so that,

\begin{eqnarray}
 M & = & WD, \\   
  &  = & s^2 r_{sys}^3 D^{-2},\\
  & = & a r_{sys}^2 .
\end{eqnarray}

\noindent Eliminating $r_{sys}$ from equations $(6)$ and $(7)$ gives;

\begin{eqnarray}
M & = & s^{-4} a^{3} D^{4},
\end{eqnarray}

\noindent 
obtained only from systemic feature accelerations and the
slope of the systemic feature rotation curve. 
That leaves equation $(5)$.
Plotting equations $(5)$ and $(8)$ in $M_{bh}-D$ space yields intersecting
curves and, if all the quantities were exactly known, an exact solution for distance
with $D= W^{1/3}s^{4/3}a^{-1}$. 

However, $a$, $s$, and $W$ are statistical quantities. If $W$ is known
rather accurately, as in our data set, then a narrow ellipse of ``allowed'' values of $M_{bh}$ and $D$ forms in  $M_{bh}-D$ space (Figure~\ref{f:outcomes}),  the 
size of which is determined by the 1$\sigma$ uncertainties in $a$, $s$, and $W$. Outside of the ellipse, solutions are not
probable as the intersecting curves separate.
The $M_{bh}-D$ joint probability distribution is highly elliptical,
which explains the high correlation coefficient \citep{Kuo2013}.

In summary,  the distance to the NGC 4258 disk can be reliably determined. Possible solutions are constrained 
to lie within an ellipse in 
$M_{bh}-D$ space, the size of which is governed by the 1$\sigma$ uncertainties in $a$, $s$, and $W$.

Note that there are other high correlations in Table~\ref{t:correlations}. However, these are caused by defining the disk warping terms
relative to $r$=0.  Had they been defined relative to $r$=$r_{mid}$, where $r_{mid}$ is the mid-radius of maser features in the disk, then
they would be small. This has no effect on the distance estimate.

\subsection{Maser Distance to NGC 4258}

Adding all of the systematic uncertainty terms in Table~\ref{t:uncertainties} yields a
distance of 7.60 $\pm$ 0.17 (random) $\pm$  0.15 (systematic) Mpc.  Combining
the random and systematic uncertainties in quadrature gives a $\pm$3\%
distance uncertainty. This is consistent with, but considerably better than,
 the result of H99.

\section{Implications of the distance for H$_{0}$} 
\label{s:discuss}

As noted in Section~\ref{s:intro}, the maser distance to NGC 4258 cannot be used to directly  
calculate a high-accuracy H$_{0}$, since it
is relatively nearby and its peculiar motion could be a large proportion of its recessional
velocity.
Instead, the role for NGC 4258 is to be an anchor for the Cepheid-calibrated
extragalactic distance scale.
Using a maser distance to NGC 4258 of 7.28 Mpc $\pm$3\% \citep{Riess2012}, \citet{Riess2011} 
obtained H$_{0,4258}$ = 74.8 $\pm$ 3.1 km s$^{-1}$ Mpc$^{-1}$. We can now revise this
value for the Hubble Constant to  H$_{0,4258}$ = 72.0 $\pm$ 3.0 km s$^{-1}$ Mpc$^{-1}$.

This is in good agreement with the $H_0$ estimated by the seven-year Wilkinson Microwave
Anistropy Probe (WMAP) data
of $71.0\pm2.5$~kms~s$^{-1}$~Mpc$^{-1}$ for standard $\Lambda$-CDM cosmology and a flat universe 
 \citep{Larson2011}, and that for WMAP$+$Baryonic Acoustic Oscillations (BAO) of 69.3$\pm$0.9 km s$^{-1}$ Mpc$^{-1}$ 
\citep{Hinshaw2012}, and in less good agreement (a $1.5\sigma$ discrepancy) with H$_0$ from the Planck satellite of 67.3$\pm$1.2 km s$^{-1}$ Mpc$^{-1}$  \citep{PlanckCollaboration2013}.
The maser cosmology project has reported two estimates of H$_0$ based on the same
procedures for estimating distance as that used here (i.e. maser distribution plus accelerations)
 with the results: UGC 3789 (distance = 49.6 Mpc)
$H_0= 68.7\pm7.1$~km~s$^{-1}$~Mpc$^{-1}$ \citep{Reid2013} and NGC 6264 (distance = 144 $\pm$ 19 Mpc) H$_0$ = 68 $\pm$ 9
 km s$^{-1}$ Mpc$^{-1}$) \citep{Kuo2013}. These masers are far enough away that H$_0$ could
be calculated  with negligible error due to the peculiar motions of the galaxies.
The weighted mean estimate of H$_0$ for these two galaxies is 68.7 $\pm$ 5.6 km s$^{-1}$ Mpc$^{-1}$.

\section{Conclusions}
\label{s:conclu}

We analyzed 18 epochs of VLBI water maser data and more than 10 years of acceleration monitoring
in Papers I and II. Here, we fitted the resulting data set using a Bayesian method to
yield a new, high-accuracy, and purely geometric maser distance to NGC 4258. 
We took particular care to assess terms of systematic uncertainty and
obtained a distance of 7.60 $\pm$ 0.23 Mpc (i.e., $\pm$ 3\%),
 which is consistent with, but much more accurate than
the H99 maser distance estimate of 7.2 $\pm$ 0.5 Mpc (7\%). 
The new distance estimate yields an H$_{0,4258}$ = 72.0 $\pm$ 3.0 km s$^{-1}$ Mpc$^{-1}$ 
which provides an important independent estimate of the Hubble Constant. 
The \citet{PlanckCollaboration2013} commented
on the ``tension" between their result and the ``local" value
of H$_0$ derived from Cepheid measurements. We note that our
result lies between these. The use of water masers to derive
the Hubble constant remains important because it does not
share the systematic uncertainties of the other methods, e.g.
the $\Lambda$-CDM model parameters in the case of the Cosmic Microwave Background derived Hubble
constant and the distance ladder assumptions associated with
the Cepheid method.
The determination of H$_0$ is approaching the level
required to impose additional constraints on the equation of state
parameter for Dark Energy \citep[][see Figure~\ref{f:h0}]{Weinberg2012}.

\acknowledgements

We thank the anonymous referee, Adam Riess and Lucas Macri for comments
that improved the manuscript.
We also thank Carolann Barrett for proofreading the manuscript.

{\it Facilities:} \facility{VLBA}, \facility{VLA}, \facility{GBT}.

\bibliographystyle{hapj}
\bibliography{humphreys_4258_for_astroph}

\begin{deluxetable}{lccc}
\tablecaption{Input Dataset \label{t:inputdata}}
\tablewidth{0pt}
\tablehead{
\colhead{} & \colhead{Reds} & \colhead{Systemics} & \colhead{Blues} 
}
\startdata
Number of data points\tablenotemark{a}                  &  151       &  187    &  32   \\
$v_{los}$ range (km s$^{-1}$)\tablenotemark{b}   &   [1227.5,1647.5]   & [382.5,577.5]   & [-516.5,-280.5]   \\
$X$ range (mas)\tablenotemark{c}   &   [7.774,2.776]   & [-0.513,0.245]   & [-4.51,-8.297]   \\
$Y$ range  (mas)\tablenotemark{c} &  [0.134,1.195]    & [-0.112,0.092]                  & [-0.014,1.060]   \\
$a_{los}$ range (km s$^{-1}$ yr$^{-1}$)&   [-0.40,0.73]& [6.96,9.81]   & [-0.72,0.04]  \\
\enddata
\tablenotetext{a}{Number of data points refers to $X$, $Y$ and $v_{los}$ data. $a_{los}$ data were not measured for each of these points.}
\tablenotetext{b}{Velocities are radio definition and LSR reference frame.}
\tablenotetext{c}{Positions are measured relative to that of maser emission at 510 km s$^{-1}$.}
\end{deluxetable}

\begin{deluxetable}{ll}
\tablecaption{High-Velocity Maser Feature Fit of \citet{Herrnstein2005} \label{t:jrhparams}}
\tablewidth{0pt}
\tablehead{
\colhead{Parameter} & \colhead{Model Value} 
}
\startdata
Distance, $D$ (Mpc)                                       & [7.2]\tablenotemark{a} \\
Black Hole Mass, $M_{bh}$ ($\times 10^7$ $M_{\odot}$)&   3.79 \\
Galaxy systemic velocity, $v_{sys}$ (km~s$^{-1}$)    & 473.5 \\
Dynamical center x-position, $X_{0}$\tablenotemark{b} (mas)           &  -0.19   \\
Dynamical center y-position, $Y_{0}$\tablenotemark{b} (mas)           &  [0.55]\tablenotemark{a} \\
Inclination, $i_{0}$ (deg)                           &  [73.80]\tablenotemark{a} \\
Inclination warp, $di/dr$ (deg/mas)                  &  1.95     \\
Position angle, $\Omega_0$ (deg)                       &  65.65     \\
Position angle warp, $d\Omega/dr$ (deg/mas)          &   5.04     \\
Position angle warp, $d^2\Omega/2dr^2$ (deg/mas$^2$) &  -0.13  \\
\enddata
\tablenotetext{a}{Values were adopted, not fitted.}
\tablenotetext{b}{Positions are measured relative to that of maser emission at 510 km s$^{-1}$.}
\end{deluxetable}

\begin{deluxetable}{lc}
\tablecaption{The Distance Base Model\label{t:basemodel}}
\tablewidth{0pt}
\tablehead{
\colhead{Parameter} & \colhead{Value\tablenotemark{a}} 
}
\startdata
Distance, $D$ (Mpc)                                                               &    7.60$\pm$0.17       \\
Black Hole Mass, $M_{bh}$ ($\times 10^7$ $M_{\odot}$)   &     4.00$\pm$0.09       \\
Galaxy systemic velocity, $v_{sys}$ (km~s$^{-1}$)              &     474.25$\pm$0.49       \\
Dynamical center x-position, $X_{0}$\tablenotemark{b} (mas)                        &      -0.204$\pm$0.005      \\ 
Dynamical center y-position, $Y_{0}$\tablenotemark{b} (mas)                         &      0.560$\pm$0.006      \\
Inclination, $i_{0}$ (deg)                                                      &    71.74$\pm$0.48        \\
Inclination warp, $di/dr$ (deg/mas)                                    &     2.49$\pm$0.11        \\
Position angle, $\Omega_0$ (deg)                                          &      65.46$\pm$0.98       \\
Position angle warp, $d\Omega/dr$ (deg/mas)                   &      5.23$\pm$0.30       \\
Position angle warp, $d^2\Omega/2dr^2$ (deg/mas$^2$) &     -0.24$\pm$0.02         \\
Eccentricity, $e$                                                                  &        0.006$\pm$0.001       \\
Periapsis angle, $\omega_0$ (deg)                                         &      293.5$\pm$64.4         \\
Periapsis angle warp, $d\omega/dr$   (deg/mas)                                  &        59.5$\pm$10.2       \\
\enddata
\tablenotetext{a}{Uncertainties have been scaled by $\sqrt{\chi^2/N}$ i.e. $\sqrt{1.403}$.}
\tablenotetext{b}{Positions are measured relative to that of maser emission at 510 km s$^{-1}$.}
\end{deluxetable}

\begin{deluxetable}{lccrc}
\tabletypesize{\scriptsize}
\tablecaption{Table of Distance Uncertainty Terms \label{t:uncertainties}}
\tablewidth{0pt}
\tablehead{
\multicolumn{1}{l}{\bf Formal Uncertainty} & \colhead{}   & \colhead{Distance} & 
\multicolumn{2}{c}{1$\sigma$ uncertainty} \\ 
\colhead{}  & \colhead{}        &     \colhead{(Mpc)}    &   \colhead{(Mpc)}  &   \colhead{(\%)}   
}
\startdata
Base Model                &                                                   &   7.596   & $\pm$0.167 &   2.20 \\
&&&\\
\tableline
&&&\\
{\bf Systematic Uncertainties}& Value in     & Distance & \multicolumn{2}{c}{$\Delta D$ from Base Model} \\ 
                              & Base Model             &          & \multicolumn{2}{c}{or rms estimate} \\
                              &              &     (Mpc)          &         (Mpc)             &        (\%)                 \\
&&&\\
\tableline
&&&\\
Different hv-feature velocity uncertainties (0.7 \& 1.3 km s$^{-1}$)    & 1.0 km s$^{-1}$               & 7.624  &   0.028   & 0.37 \\
Different y-error floor (20 $\mu$as)                           & 30 $\mu$as         & 7.639  & 0.043 & 0.57 \\
Different acceleration error floor  (0.5 km s$^{-1}$yr$^{-1}$)   & 0.3 km s$^{-1}$yr$^{-1}$                                    & 7.564   &  -0.032 & 0.42\\
Different initial conditions\tablenotemark{a}                   & ---         &  ---      & $\pm$0.114   & 1.50 \\
Assuming eccentricity is zero                                   &  ---        &  7.619    & 0.023 & 0.30   \\
Inclusion of $d^2i/dr^2$ term (solves to -0.112$\pm$0.034 deg mas$^{-2}$)                              &  ---                 & 7.562    &  -0.034&  0.45  \\
Unmodeled spiral structure                                     & ---            &   ---           & $\pm$0.076     &  1.00    \\
&&&\\
\tableline
&&&\\
\multicolumn{3}{l}{{\bf Systematic Uncertainties Added in Quadrature}} &   0.15       &   2.0 \%     \\
&&&\\
\tableline
&&&\\
 \multicolumn{3}{l}{{\bf Uncertainty in the Maser Distance to NGC 4258\tablenotemark{b} }} & 0.23      &   3.0  \%     \\
&&&\\
\enddata
\tablenotetext{a}{This includes different seeds in the random number generator, and different initial distances in the model.}
\tablenotetext{b}{The final uncertainty is
calculated by adding individual sources of uncertainty in quadrature.}
\end{deluxetable}

\begin{deluxetable}{lll} 
\tablewidth{0pt}
\tablecaption{Unmodeled Spiral Structure Test
\label{t:spiralstructure}}
\tablehead{Parameter & Model Value& Fitted Value}
\startdata
Distance, $D$ (Mpc)                                    & 7.20   &  7.46     \\
Black Hole Mass, $M_{bh}$ ($\times 10^7$ $M_{\odot}$)& 3.80   &   3.96     \\
Galaxy systemic velocity, $v_{sys}$ (km~s$^{-1}$)    & 474.00 &  436.20  \\
Dynamical center x-position, $X_{0}$ (mas)           & 0.000   & 0.17   \\
Dynamical center y-position, $Y_{0}$ (mas)           & 0.000   & 0.00    \\
Inclination, $i_0$ (deg)                           & 90.0  &  89.97  \\
Position angle, $\Omega_0$ (deg)                       & 90.0  &  90.00       \\
Eccentricity, $e$                                     & 0.00  &  0.078       \\
Periapsis angle, $\omega_0$ (deg)                     & 0.00  &  -33.13       \\
Periapsis angle gradient, $d\omega/dr$ (deg/mas)      & 0.00  & 13.57 \\
\enddata
\end{deluxetable}

\begin{deluxetable}{lrrrrrrrrrrrrr} 
\tabletypesize{\scriptsize}
\tablewidth{0pt}
\tablecaption{Parameter Correlation Matrix\label{t:correlations} (see Section~\ref{ss:correlation})}
\tablehead{
Parameter & \multicolumn{1}{c}{$D$}  & \multicolumn{1}{c}{$M_{bh}$}  &\multicolumn{1}{c}{$v_{sys}$}& \multicolumn{1}{c}{$x_{0}$} & \multicolumn{1}{c}{$y_{0}$} &  \multicolumn{1}{c}{$i_{0}$} &  $di/dr$ & \multicolumn{1}{c}{$\Omega$} & \multicolumn{1}{c}{$d\Omega/dr$} & \multicolumn{1}{r}{$d^2\Omega/2dr^2$} &  \multicolumn{1}{c}{$e$}   &  \multicolumn{1}{c}{$\omega$} &  \multicolumn{1}{c}{$d\omega/dr$} 
}
\startdata
   $D$      & 1.000  & 0.998 & 0.011 &-0.009 & 0.030 &-0.089 & 0.108 &  0.094 &-0.090 & 0.089 &-0.060 & 0.301& -0.285 \\
 $M_{bh}$    & 0.998  & 1.000 &-0.023 &-0.026 & 0.025 &-0.083 & 0.102 &  0.098 &-0.095 & 0.095 &-0.058 & 0.311& -0.298\\
 $v_{sys}$   & 0.011  &-0.023 & 1.000 & 0.062 & 0.210 &-0.124 & 0.119 & -0.216 & 0.240 &-0.239 & 0.051 &-0.413&  0.423\\
$X_{0}$      &-0.009  &-0.026 & 0.062 & 1.000 &-0.115 & 0.192 &-0.167 &  0.052 &-0.047 & 0.045 &-0.695 &-0.241&  0.159\\
$Y_{0}$      & 0.030  &0.025  &0.210  &-0.115 & 1.000 &-0.416 & 0.253 &  0.014 & 0.068 &-0.099 & 0.110 &-0.021&  0.031\\
 $i_{0}$ &-0.089  &-0.083 &-0.124 & 0.192 &-0.416 & 1.000 &-0.979 &  0.073 &-0.099 & 0.103 &-0.167 & 0.083& -0.095\\
 $di/dr$ & 0.108  &0.102  &0.119  &-0.167 & 0.253 &-0.979 & 1.000 & -0.075 & 0.090 &-0.088 & 0.147 &-0.086&  0.097\\
 $\Omega_0$ & 0.094  &0.098  &-0.216 & 0.052 & 0.014 & 0.073 &-0.075 & 1.000 &-0.990 & 0.972  &0.047  &0.211& -0.199\\
$d\Omega/dr$ &-0.090  &-0.095 & 0.240 &-0.047 & 0.068 &-0.099 & 0.090 & -0.990 & 1.000 &-0.995 &-0.037 &-0.214&  0.200\\
$d^2\Omega/2dr^2$ & 0.089  & 0.095 &-0.239 & 0.045 &-0.099 & 0.103 &-0.088 & 0.972 &-0.995 & 1.000  &0.028  &0.211&-0.197\\
 $e$     &-0.060  &-0.058 & 0.051 &-0.695 & 0.110 &-0.167 & 0.147 & 0.047 &-0.037 & 0.028  &1.000  &0.074& 0.015\\
$\omega_0$  & 0.301  &0.311  &-0.413 &-0.241 &-0.021 & 0.083 &-0.086 & 0.211 &-0.214 & 0.211  &0.074  &1.000&-0.979\\
$d\omega/dr$  &-0.285  &-0.298 & 0.423 & 0.159 & 0.031 &-0.095 & 0.097 &-0.199 & 0.200 &-0.197  &0.015 &-0.979& 1.000\\
\enddata
\end{deluxetable}

\clearpage

\begin{figure}
\includegraphics[angle=0,scale=0.7]{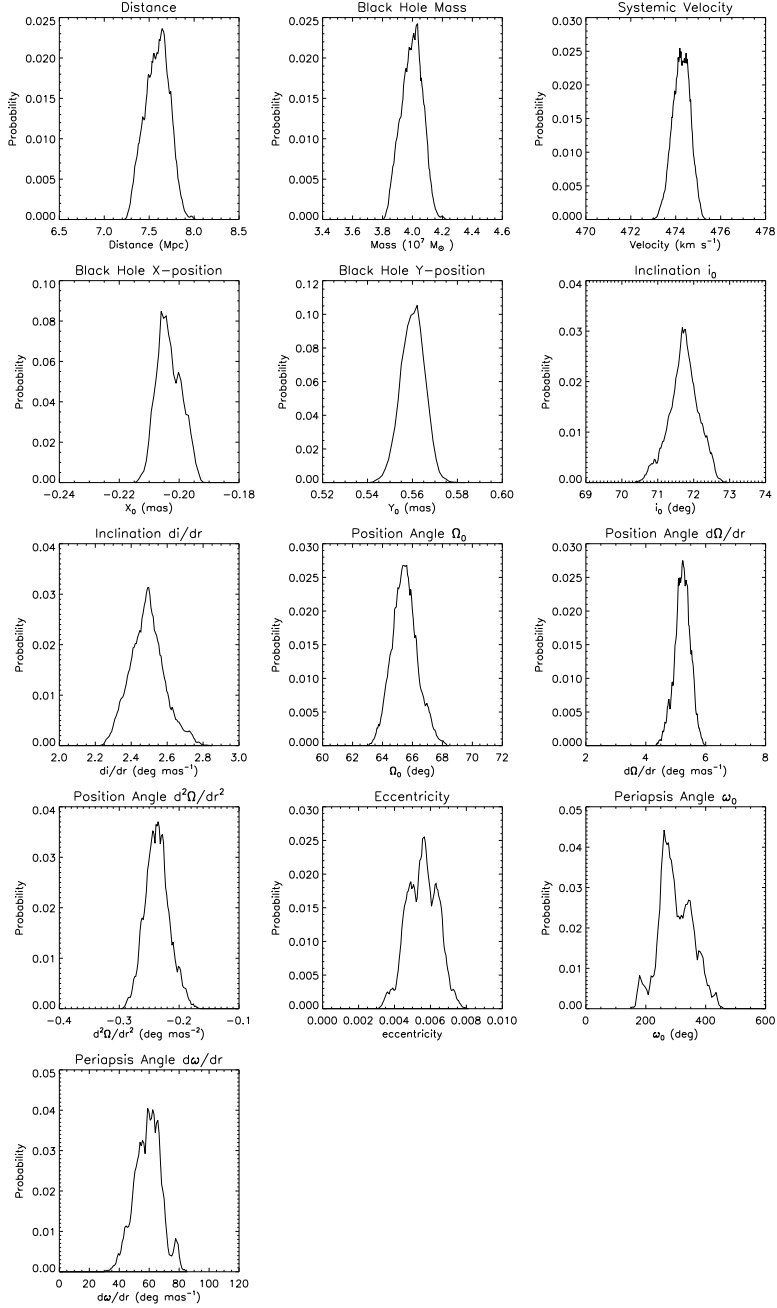}
\vspace{-0.5cm}
\caption{Marginalized posteriori probability distributions for the global disk parameters describing the distance base model.
These have been produced by binning 10,000,000 MCMC  trial values for each parameter.
\label{f:circmodel_pdfs_1}}
\end{figure}
\clearpage

\begin{figure}
\includegraphics[scale=0.7]{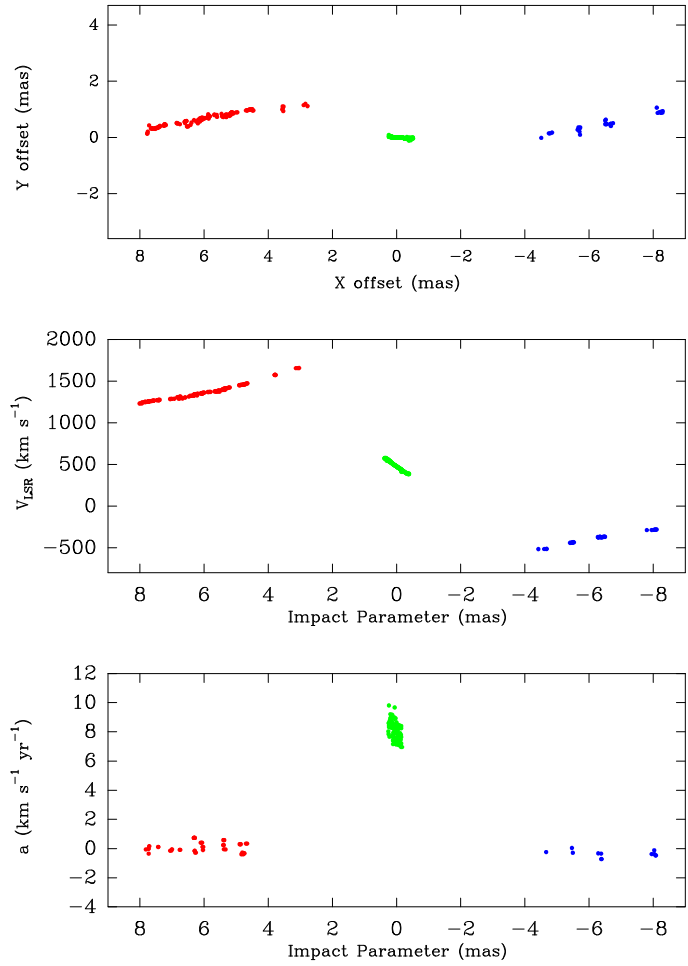}
\vspace{-0.5cm}
\caption{Input data to the model: {\it (top)} maser sky positions; {\it (middle)} P-V diagram; and,  {\it (bottom)} accelerations.
\label{f:inputdata}}
\end{figure}

\clearpage

\begin{figure}
\includegraphics[scale=0.7]{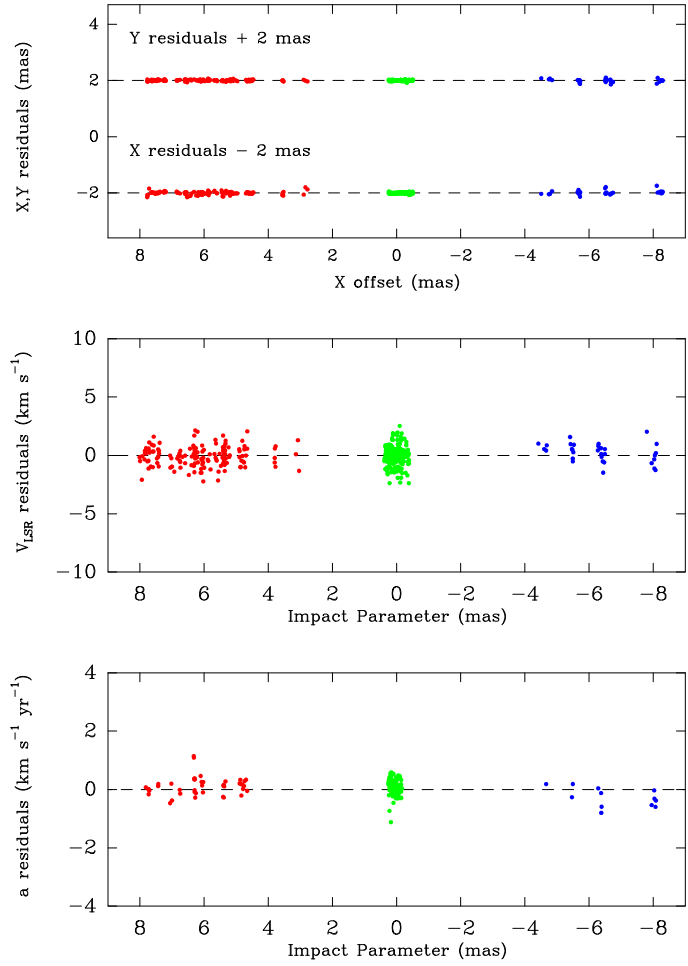}
\vspace{-0.5cm}
\caption{Residuals from the base model fit to the data: {\it (top)} maser sky position residuals, offset by $\pm$2 mas for clarity;  {\it(middle)} velocity residuals; and, {\it(bottom)} acceleration residuals. 
\label{f:residuals}}
\end{figure}

\clearpage

\begin{figure}
\includegraphics[angle=0,scale=0.7]{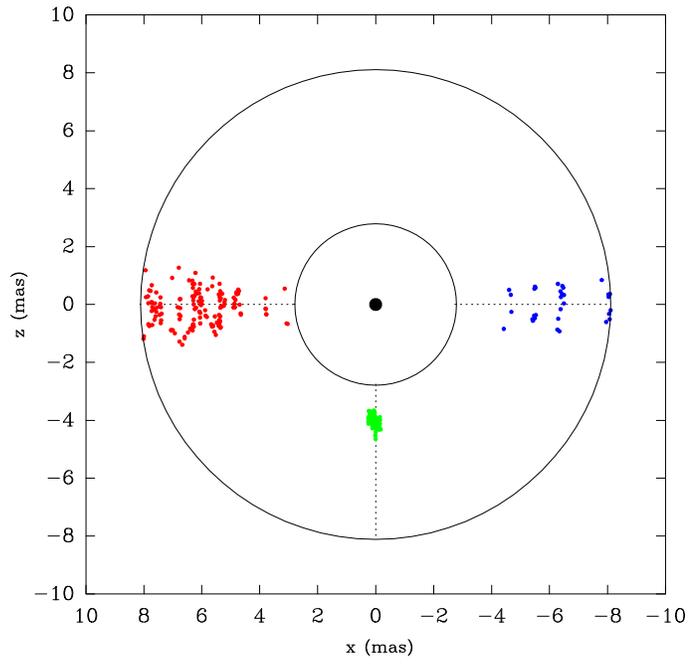}
\vspace{-2.0cm}
\caption{View of the disk plane for the base model. Deprojected maser positions are shown in red, green and blue for the red-shifted, 
systemic, and blue-shifted masers, respectively. 
} 
\end{figure}

\clearpage

\begin{figure}
\includegraphics[angle=0,scale=0.7]{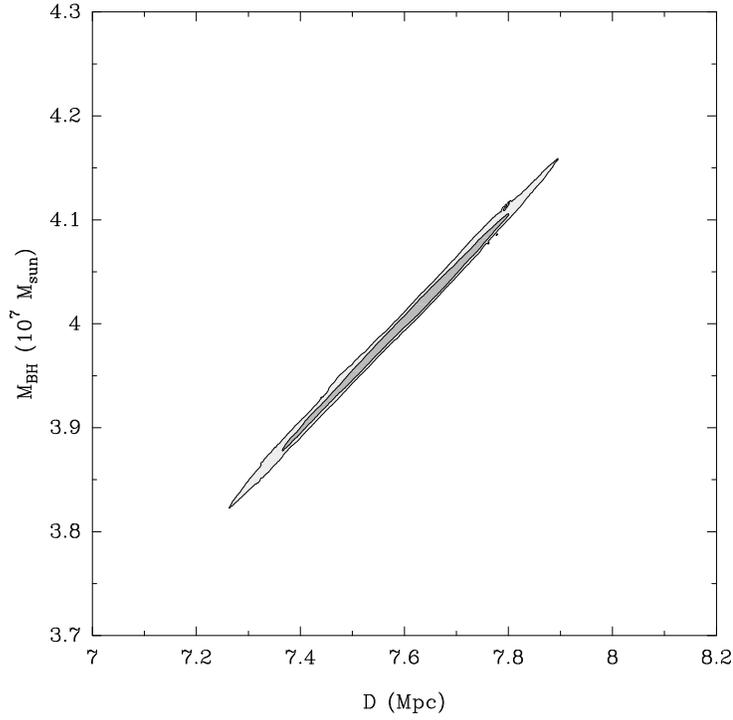}
\vspace{-2cm}
\caption{2D histogram of distance and mass outcomes from the
10,000,000 MCMC trials. Bin sizes used to compute the
plot were 0.001 Mpc and 0.001 M$_{\odot}$ respectively. Greyscale contours
code the 68\% and 95\% confidence intervals.  
\label{f:outcomes}} 
\end{figure}

\clearpage

\begin{figure}
\includegraphics[angle=0,scale=0.7]{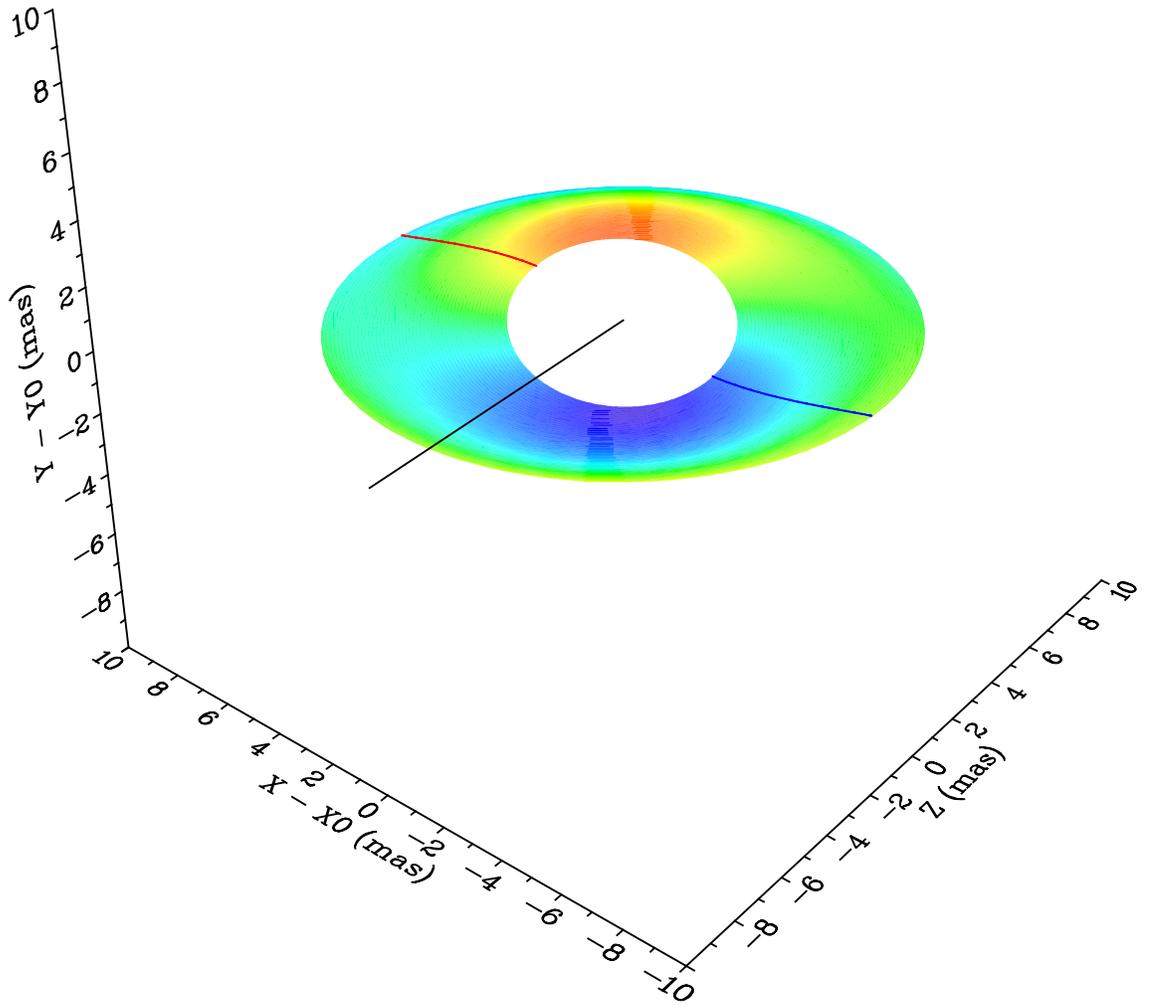}
\vspace{-2cm}
\caption{Basemodel maser disk seen from a viewpoint at [$-$40,40,$-$50] mas. The LOS is shown as a line extending
 beyond the outer edge of the disk 
in black along the Z-direction. Solid color contours show disk elevation, where red is the maximum and dark blue is the minimum. 
The disk midline is shown for redshifted emission (red line) and blueshifted emission (blue line), respectively. Low-velocity (systemic) 
masers lie in the concavity on the front side of the disk.
\label{f:3dwarp}} 
\end{figure}

\clearpage

\begin{figure}
\includegraphics[angle=0,scale=0.7]{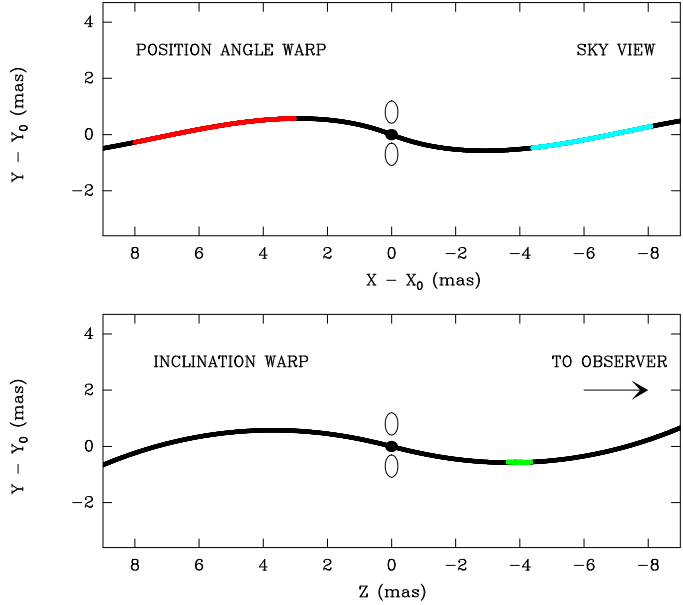}
\vspace{-4cm}
\caption{Cross-sections of the warped disk of the basemodel.
{\it (top)} cut of the disk along the disk midline, showing the
position angle warp given by 
$\Omega_r = \Omega_0 + r d\Omega/dr + r^2 d^2\Omega/dr^2$. 
Total extents of  red- and blue-shifted maser regions from the basemodel are overplotted.
{\it (bottom)} cut of the disk along the line-of-sight to the black hole,
showing the inclination warp given by $i_r = i_0 + r di/dr$. Total extent of the systemic maser
region (for systemic masers with acceleration measurements) from the basemodel is overplotted.
 {\it (both panels)} the northern oval represents the observed position of northern radio 
jet emission \citep{Herrnstein1997} and the southern oval is a reflection of the northern one through the disk center \citep[following][]{Herrnstein2005}.
\label{f:crosssections}} 
\end{figure}

\clearpage

\begin{figure}
\includegraphics[angle=0,scale=0.7]{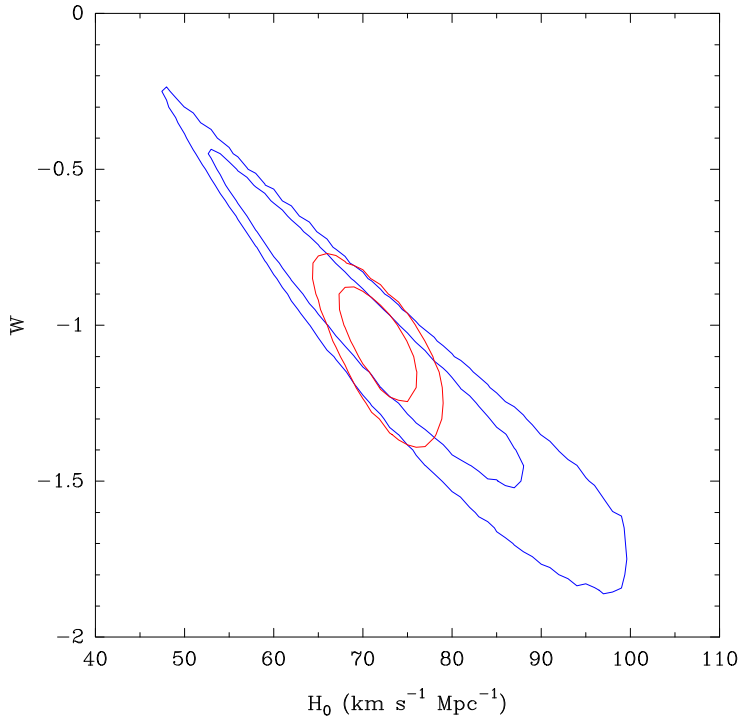}
\vspace{-2cm}
\caption{Constraint on Dark Energy. 2-D probability density functions (PDFs) for 
a constant equation of state of dark energy ($w$) 
and H$_{0}$ with 95\% and 68\% confidence contours. {\it (Blue contours)} Blue 
contours were generated 
by binning the parameter values from Markov chains 
from the WMAP 7-year data (modeled with a constant-$w$, $\Lambda$CDM model that incorporates the 
effects of the SZ effect and gravitational lensing).  
{\it (Red contours)} Red contours combine the WMAP PDF and the constraint
that H$_{0}$=72.0$\pm$3.0 km s$^{-1}$ Mpc$^{-1}$ from the results presented in this
paper.  
Fitting a Gaussian to the marginalized 1-dimensional
PDF for $w$ yields $w=-1.06\pm0.12$ ($\pm68$\% confidence).   
\label{f:h0}} 
\end{figure}

\clearpage

\begin{figure}
\hspace{-2cm}
\includegraphics[angle=270,scale=0.5]{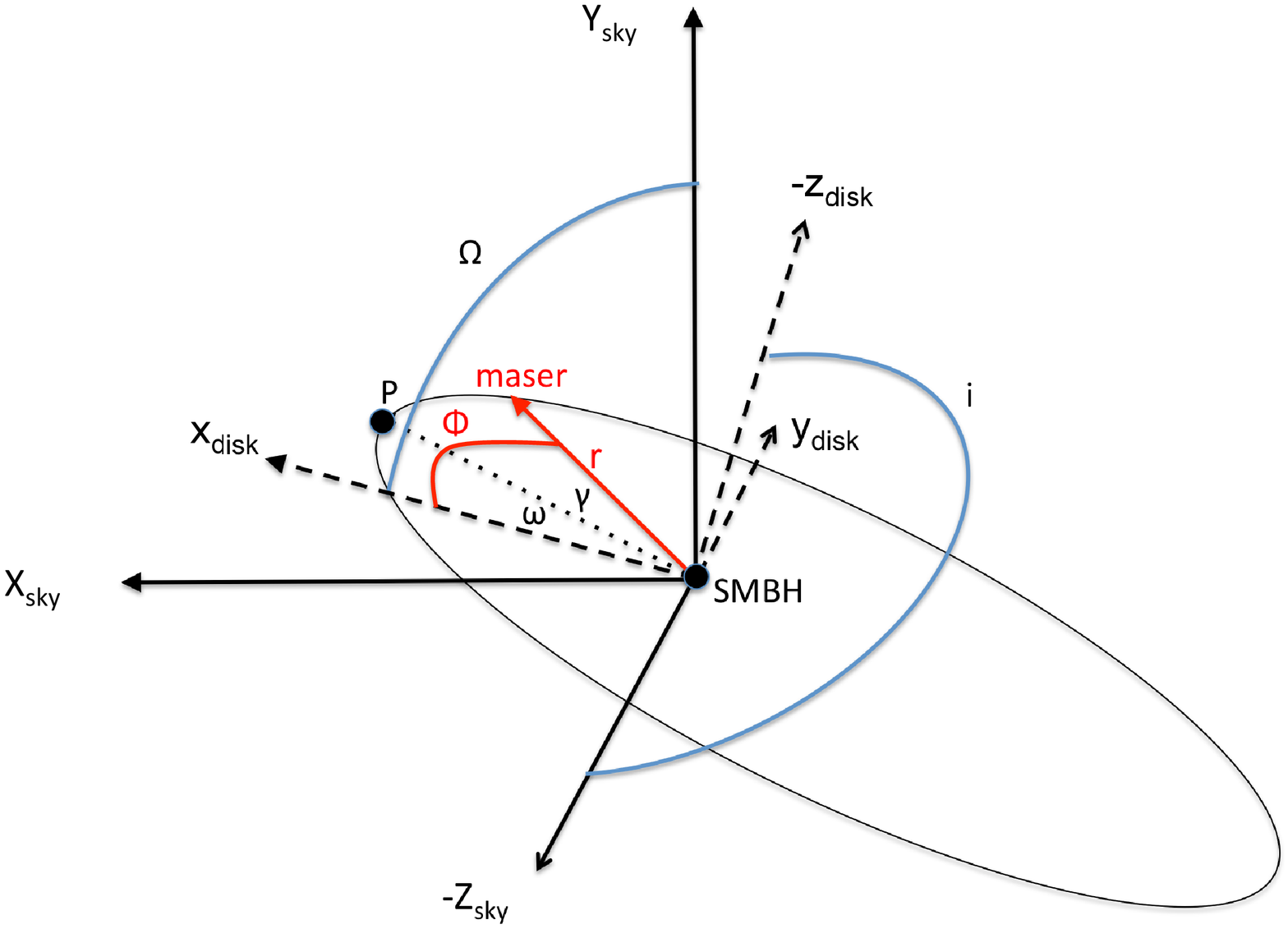}
\caption{Geometry of the maser disk - sky plane coordinate systems.
$X_{sky}$ - Y$_{sky}$ is the sky plane and $x_{disk}$ - y$_{disk}$ is
the maser disk plane. The supermassive black hole (SMBH) marks the
zero-point of both coordinate systems. $-z_{disk}$ marks the
direction of the negative angular momentum for the maser disk, and
the angle between  $-z_{disk}$ and $-Z_{sky}$ which is the direction
to the observer is the inclination angle $i$. Perihelion of the maser
orbit displayed is marked with a $P$. The position of maser features
in the disk is measured from the disk $x$-axis, which is confined to
lie in the sky plane,  and is the angle $\phi$ for each feature.
The periapsis angle is the angle $\omega$ and the disk
position angle, $\Omega$, is measured in the
sky-plane in the direction East of North to the disk $x$-axis.
\label{f:geometry}} 
\end{figure}

\appendix

\section{APPENDIX A: Coordinate Systems} 
\label{a:coordinates}

Consider a maser disk plane coordinate system  $(x,y,z)$ and a sky-plane system denoted  $(X,Y,Z)$
with a super-massive black hole at the common origin of both systems (Figure~\ref{f:geometry}).

Drawing on the notation of \citet{Peiris2003}, the sky-plane system has the $(X,Y)$ plane as the sky plane with
positive $X$ pointing east and positive  $Y$ pointing north, a right-handed coordinate
system. The positive $Z$-axis points along the line-of-sight
away from the observer, and the line-of-sight velocity 
$v_{los} = \dot{Z}$ is positive for receding objects.
The second system is the disk-plane system in which coordinates of the disk plane are
denoted $(x,y,z)$. The $(x,y)$ plane is in the plane of the disk with the
positive $x$ axis pointing along the sky $X$-axis.
The positive $z$ axis points in the direction of the disk angular 
momentum vector (i.e., also a right-handed coordinate system).

To convert between
sky $(X,Y,Z)$ and disk $(x,y,z)$ coordinate frames, for a flat, inclined maser disk 
we perform rotations about two Euler angles ($\Omega$, $i$)
where $\Omega$ is the angle to the disk $x$-axis measured East
of North (the $Y$-axis) in the $(X, Y)$ plane and is the disk position angle and $i$ is the inclination angle
measured between $-Z$ and $-z$.   The first rotation is by $i$ about the sky $X$-axis and the second rotation is by $\Omega$ about the sky $Z$-axis, i.e.,

\begin{equation} 
\nonumber
\pmatrix{X \cr Y \cr Z \cr} = \pmatrix{\sin\Omega & -\cos\Omega & 0 \cr
\cos\Omega & \sin\Omega & 0 \cr
0 & 0 & 1 }\pmatrix{1 & 0 & 0 \cr 0 & \cos i & -\sin i \cr 0 & \sin i & \cos i }\pmatrix{x \cr y \cr z \cr} , 
\nonumber
\end{equation}

\noindent where the coordinates of a maser feature in the disk frame are $(x, y, z)=(r\cos\phi,r\sin\phi,0)$.

The coordinates of masers in the sky reference frame are therefore given by:

\begin{eqnarray}
X & = & r(\sin\Omega\cos\phi - \cos\Omega\cos i \sin\phi) \nonumber\\
Y & = & r(\cos\Omega\cos\phi + \sin\Omega\cos i \sin\phi)\\
Z & = & r\sin i\sin\phi. \nonumber
\end{eqnarray}

\section{APPENDIX B: Velocity \& Acceleration}
\label{a:velocity}

For defining velocity, the angle of the maser in the disk with respect to perihelion, 
periapsis angle $\omega$, now becomes relevant i.e., $\gamma=(\phi -\omega)$.
Velocity in disk polar coordinates is given by 
$v=\dot{r} +r\dot{\gamma}$,  where
$v_r=\dot{r}$ and $v_{\gamma}=r\dot{\gamma}$ are radial and 
tangential velocity components, respectively. In this Appendix, $r$ is a linear radius
rather than an angular one.
Specific angular momentum $h$ is given by 
$h   =  2 \pi a b/P $, where  $a$ and $b$ are the semi major and minor axes, 
respectively, and $P$ is the orbital period. As
$b=a \sqrt{(1 - e^2)}$ and $P^2=4\pi^2a^3/GM$, $h   =  \sqrt{a(1-e^2)GM}$. 
Since $\dot{\gamma} =  h/r^2$ and 
 $r=a(1-e^2)/(1+e\cos\gamma)$, $v_{\gamma}  =  r\dot{\gamma}  = rh / r^2 
  = \sqrt{a(1-e^2)GMr^2/r^4} 
          =  \sqrt{GM(1+e\cos\gamma)/r}$,  and 
 $v_{r}  =  \dot{r} 
        = a(1-e^2)/(1+e\cos\gamma)^2) e\sin\phi \dot{\gamma} 
        =  \sqrt{GM/r(1+e\cos\gamma)} e\sin\gamma$.

Velocity components in the disk frame are denoted by 
$(v_x,v_y,v_z)=
(v_r\cos\phi - v_{\gamma}\sin\phi,v_r\sin\phi+v_{\gamma}\cos\phi,0)$ 
and in the sky frame by, 

\begin{eqnarray}
v_X &=& v_r(\sin\alpha\cos\phi-\cos\alpha\cos i\sin\phi) \nonumber\\
    & & + v_{\gamma}(\cos\alpha\cos i\cos\phi - \sin\alpha\sin\phi) \nonumber\\
v_Y &=& v_r(\cos\alpha\cos\phi+\sin\alpha\cos i\sin\phi) \nonumber\\
    & & + v_{\gamma}(\sin\alpha\cos i\cos\phi-\cos\alpha\sin\phi)\\
v_Z &=& v_r\sin i\sin\phi+v_{\gamma}\sin i\cos\phi. \nonumber
\end{eqnarray}

Acceleration is given by $a= a_r + a_{\gamma} = \ddot{r} - r (\dot{\gamma})^2 + r \ddot{\gamma} + 2 \dot{r} \dot{\gamma}$, where the centripetal acceleration component $a_r = -GMr^{-2}$ and  $a_{\gamma} = 0$. In the disk frame, the components of $a_r$ are $(a_{x},a_{y},a_{z})=(-a_r\cos\phi,-a_r\sin\phi,0)$, so that in the line-of-sight to the observer, $a_Z = -a_r\cos i\sin i\sin \phi - a_r\sin\phi\sin i\cos\phi  
 = -GMr^{-2}\sin i\sin\phi$.

\section{APPENDIX C: Relativistic Effects}
\label{a:relativity}

When modeling water maser emission from AGN disks, one must
relate a model velocity to the observed (e.g., optical or radio-definition) velocities. 

Starting in a reference frame located at the focus of the maser orbits (the black-hole frame),
gas clouds at radius, $r$, orbit with velocity, $v_{orb} = \sqrt{GM/rD}$, where $M$ is the mass
of the black hole (and providing one is not near the strong gravity regime). Our
observing frame is receding from the black-hole frame at velocity $v_{rec}$.

We will use the relativistic Doppler equation \citep[see][p.111,
eq. 4.11]{Rybicki1986}, which gives the relation between frequency, $f$ , in the source rest frame
to the frequency, $F$ , in an observers frame moving with velocity $v$ relative to the
source:

\begin{equation}
F = f \Gamma^{-1}  / (1 - \frac{v}{c} \cos \theta), \\
\end{equation}

\noindent where $\theta$ is the angle between the vector $v$ and our line of sight and

\begin{equation}
\Gamma = 1/\sqrt{1 - (v^2/c^2)} .
\end{equation}

\noindent (Note, $v$ is implicitly $\ge$ 0 and $\cos\theta$ can be positive or negative, i.e., coming toward
or away from the observer, respectively).

In a frame at rest with a masing cloud, the emission is at the rest frequency
$f_0$. Such a cloud is moving with respect to the black-hole frame by its orbital
velocity, $v_{orb}$, which can be decomposed into parallel (toward the observer) and perpendicular components, $v_{\parallel}$ and v$_{\perp}$ . 
Imagine observing in this frame at the distance
of the Sun, but still at rest in the black-hole frame (i.e., zero recessional velocity).
Using the relativistic Doppler equation,

\begin{equation}
f = f_0 \Gamma^{-1}/(1 + \frac{v_{\parallel}}{c}). 
\end{equation}

\noindent Next, transform from the black-hole frame to the observers frame, which is receding at 
velocity $v_{rec}$ along the line of sight. Again using the relativistic Doppler
equation:

\begin{equation}
F = f \Gamma^{-1}/(1 + \frac{v_{rec}}{c}) .
\end{equation}

\noindent $F$ gives observed frequency accounting for all special relativistic effects, including time 
dilation and light-travel effects. However, as the photons travel away from
the black hole, they experience a general relativistic ``gravitational redshift". This
reduces the observed frequency further such that the observed frequency is given
by

\begin{equation}
F_{gr} = F \sqrt{1 - R_{sch}/r}, 
\end{equation}

\noindent where $R_{sch} = 2GM/c^2$.

Finally, if, for example,  the observational data we seek to model uses the optical
definition of velocity

\begin{equation}
v'_{los} \equiv c (\frac{f_0}{F_{gr}} -1),
\end{equation}

\noindent or equivalent conversions for other velocity definitions can be made.

\end{document}